\documentstyle[12pt,thmsa,sw20lart]{article}

\input tcilatex
\QQQ{Language}{
British English
}

\begin{document}

\section{Introduction}

Intrinsic spin and orbital angular momentum are fundamentally the same. One
difference between them appears to be that while orbital angular momentum
has a wave-mechanical description through Legendre's equation, spin does
not. However, in this paper, we propose a differential eigenvalue equation
for spin 1/2.

The arguments leading to our proposed equation are based on the Land\'e
interpretation of quantum mechanics[1-4]. In this interpretation, the
probability-amplitude character of an eigenfunction or wave function is
stressed. Such a probability amplitude connects well-defined initial and
final states[5]. Perhaps the most important such probability amplitudes
result from solution of the time-independent Schr\"odinger equation

\begin{equation}
-\frac{\hbar ^2}{2m}\nabla ^2\psi ({\bf r})+V({\bf r})\psi ({\bf r})=E\psi (%
{\bf r}).  \label{on1}
\end{equation}
The eigenfunction $\psi ({\bf r})$ connects an initial state defined by the
eigenvalue $E$ and a final state defined by the eigenvalue ${\bf r}$. To
emphasize the initial and final state, we write the eigenfunction as $\psi (%
{\bf r})=\psi (E_n;{\bf r})$, so that the equation reads 
\begin{equation}
-\frac{\hbar ^2}{2m}\nabla ^2\psi (E_n;{\bf r})+V({\bf r})\psi (E_n;{\bf r}%
)=E_n\psi (E_n;{\bf r}).  \label{tw2}
\end{equation}
The Schr\"odinger equation is, of course, a typical eigenvalue equation.
Another example is the Legendre equation

\begin{equation}
L^2(\theta ,\varphi )Y_{lm}(\theta ,\varphi )=l(l+1)\hbar ^2Y_{lm}(\theta
,\varphi ).  \label{th3}
\end{equation}
In this equation, the initial state corresponds to the eigenvalues $%
l(l+1)\hbar ^2$ and $m\hbar $ while the final state corresponds to the
angular position $(\theta ,\varphi )$[6]. For this reason, we may write the
eigenfunctions as $Y_{lm}(\theta ,\varphi )=Y(l,m;\theta ,\varphi ).$

We notice that in both the Schr\"odinger and the Legendre equations, the
operator is cast in terms of the variables which also define the final
eigenvalue. We take this to be a general feature, which we exploit in our
search for a differential equation for spin.

\section{Spin Eigenvectors}

As conventionally treated, spin is such that both the initial and final
states correspond to discrete eigenvalues. A differential eigenvalue
equation requires a continuous variable for the operator to act on. But, the
Pauli spin vectors 
\begin{equation}
\lbrack \xi _{z,z}^{+}]=\left( 
\begin{array}{c}
1 \\ 
0
\end{array}
\right) \;\;\;\;\;\text{and\ \ \ \ \ \ }[\xi _{z,z}^{-}]=\left( 
\begin{array}{c}
0 \\ 
1
\end{array}
\right)  \label{fo4}
\end{equation}
do not appear to offer a continuous variable which might make possible a
differential treatment of spin. However, the Pauli vectors are specialised
forms of the generalized spin eigenvectors[5]

\begin{equation}
\lbrack \xi _{\widehat{{\bf a}},\widehat{{\bf b}}}^{+}]=\left( 
\begin{array}{c}
\chi ((+\frac 12)^{(\widehat{{\bf a}})};(+\frac 12)^{(\widehat{{\bf b}})})
\\ 
\chi ((+\frac 12)^{(\widehat{{\bf a}})};(-\frac 12)^{(\widehat{{\bf b}})})
\end{array}
\right) =\left( 
\begin{array}{c}
\cos \theta /2\cos \theta ^{\prime }/2+e^{i(\varphi -\varphi ^{\prime
})}\sin \theta /2\sin \theta ^{\prime }/2 \\ 
\cos \theta /2\sin \theta ^{\prime }/2-e^{i(\varphi -\varphi ^{\prime
})}\sin \theta /2\cos \theta ^{\prime }/2
\end{array}
\right)  \label{si6}
\end{equation}
and

\begin{equation}
\lbrack \xi _{\widehat{{\bf a}},\widehat{{\bf b}}}^{-}]=\left( 
\begin{array}{c}
\chi ((-\frac 12)^{(\widehat{{\bf a}})};(+\frac 12)^{(\widehat{{\bf b}})})
\\ 
\chi ((-\frac 12)^{(\widehat{{\bf a}})};(-\frac 12)^{(\widehat{{\bf b}})})
\end{array}
\right) =\left( 
\begin{array}{c}
\sin \theta /2\cos \theta ^{\prime }/2-e^{i(\varphi -\varphi ^{\prime
})}\cos \theta /2\sin \theta ^{\prime }/2 \\ 
\sin \theta /2\sin \theta ^{\prime }/2+e^{i(\varphi -\varphi ^{\prime
})}\cos \theta /2\cos \theta ^{\prime }/2
\end{array}
\right) .  \label{se7}
\end{equation}
They result when we set all the angles equal to zero.

The elements of the eigenvectors Eqns. (\ref{si6}) and (\ref{se7}) are
probability amplitudes $\chi ((m_i)^{(\widehat{{\bf a}})};(m_f)^{(\widehat{%
{\bf b}})})$ for spin projection measurements from the direction $\widehat{%
{\bf a}}$ (polar angles ($\theta ^{\prime },\varphi ^{\prime }$)) to the
direction $\widehat{{\bf b}}$ (polar angles ($\theta ,\varphi $))[5]. The
projection quantum numbers $m_i$ and $m_f$ give the spin projections
corresponding to the appropriate directions. This probability amplitude
represents the probability of obtaining the spin projection $m_f\hbar $
along the direction $\widehat{{\bf b}}$ if the spin projection is initially $%
m_i\hbar $ along the direction $\widehat{{\bf a}}$. Our notation for the
eigenvectors takes cognisance of the quantization directions that define the
probability amplitudes that constitute the elements of the eigenvectors.
This is why the Pauli spin vectors Eqns. ($\ref{fo4}$) contain the indices $z
$ and $z$. They contain as their elements probability amplitudes for spin
projection measurements from the $z$ direction to the $z$ direction.

The most important aspect of the eigenvectors Eqns. (\ref{si6}) and (\ref
{se7}) is that the polar angles $(\theta ,\varphi )$ of the final
quantization direction $\widehat{{\bf b}}$ give us continuous variables on
which a differential operator may act. In a preliminary search for a
differential operator for spin, we may simplify our task by setting the
angles referring to the initial eigenvalue to fixed values, so that we have
only the angles for the final observable as continuous variables. It is
easiest to set the polar angles of $\widehat{{\bf a}}$ to $(\theta ^{\prime
},\varphi ^{\prime })=(0,0),$ so that $\widehat{{\bf a}}$ defines the $z$
axis. In that case, we obtain

\begin{equation}
\lbrack \xi _{z,\widehat{{\bf b}}}^{+}]=\left( 
\begin{array}{c}
\chi ((+\frac 12)^{(\widehat{{\bf k}})};(+\frac 12)^{(\widehat{{\bf b}})})
\\ 
\chi ((+\frac 12)^{(\widehat{{\bf k}})};(-\frac 12)^{(\widehat{{\bf b}})})
\end{array}
\right) =\left( 
\begin{array}{c}
\cos \theta /2 \\ 
-e^{i\varphi }\sin \theta /2
\end{array}
\right) \;\;\;\;\;  \label{ei8}
\end{equation}
and 
\begin{equation}
\text{\ }[\xi _{z,\widehat{\QTR{textbf}{b}}}^{-}]=\left( 
\begin{array}{c}
\chi ((-\frac 12)^{(\widehat{{\bf k}})};(+\frac 12)^{(\widehat{{\bf b}})})
\\ 
\chi ((-\frac 12)^{(\widehat{{\bf k}})};(-\frac 12)^{(\widehat{{\bf b}})})
\end{array}
\right) =\left( 
\begin{array}{c}
\sin \theta /2 \\ 
e^{i\varphi }\cos \theta /2
\end{array}
\right) ,  \label{ei8a}
\end{equation}
which we write in the more symmetrical forms

\begin{equation}
\lbrack \xi _{z,\widehat{{\bf b}}}^{+}]=\left( 
\begin{array}{c}
e^{-i\varphi /2}\cos \theta /2 \\ 
-e^{i\varphi /2}\sin \theta /2
\end{array}
\right) \;\;\;\;\;\text{and\ \ \ \ \ \ }[\xi _{z,\widehat{\QTR{textbf}{b}}%
}^{-}]=\left( 
\begin{array}{c}
e^{-i\varphi /2}\sin \theta /2 \\ 
e^{i\varphi /2}\cos \theta /2
\end{array}
\right) .  \label{ni9}
\end{equation}

\section{Differential Operators}

We seek a differential operator $[S_{z,\widehat{{\bf b}}}]$ such that

\begin{equation}
\lbrack S_{z,\widehat{{\bf b}}}][\xi _{z,\widehat{{\bf b}}}^{\pm }]=\pm
\frac 12\hbar [\xi _{z,\widehat{{\bf b}}}^{\pm }].  \label{te10}
\end{equation}
The notation for the operator again takes into account the initial and final
quantization directions contained in the eigenvectors of the operator. Since
the solutions are $2\times 1$ matrices, of which there are two independent
ones, the operator must be a $2\times 2$ matrix, with elements that are
differential operators. We therefore adopt the form

\begin{equation}
\lbrack S_{z,\widehat{{\bf b}}}]=\left( 
\begin{array}{cc}
A & B \\ 
C & D
\end{array}
\right)  \label{el11}
\end{equation}
for the operator. As the operator is analogous to $L_z$ in the theory of
orbital angular momentum, we assume that the differential operators are of
first degree in the variables. Therefore, the elements are taken to have the
structure

\begin{equation}
A=a_1\frac \partial {\partial \theta }+a_2\frac \partial {\partial \varphi },
\label{tw12}
\end{equation}
\begin{equation}
B=b_1\frac \partial {\partial \theta }+b_2\frac \partial {\partial \varphi },
\label{th13}
\end{equation}

\begin{equation}
C=c_1\frac \partial {\partial \theta }+c_2\frac \partial {\partial \varphi }
\label{fo14}
\end{equation}
and

\begin{equation}
D=d_1\frac \partial {\partial \theta }+d_2\frac \partial {\partial \varphi },
\label{fi15}
\end{equation}
with $a_1$, $a_2$ .... $d_2$ to be determined from the condition Eqn. (\ref
{te10}).

By substituting into the eigenvalue equation, we find

\begin{equation}
b_1=b_2=c_1=c_2=0,  \label{se17}
\end{equation}
\begin{equation}
a_1=-\hbar \sin \theta \text{, }a_2=i\hbar \cos \theta  \label{ei18}
\end{equation}
and 
\begin{equation}
d_1=\hbar \sin \theta ,\text{ }d_2=i\hbar \cos \theta .  \label{ni19}
\end{equation}
Hence

\begin{equation}
\lbrack S_{z,\widehat{{\bf b}}}]=\hbar \left( 
\begin{array}{cc}
-\sin \theta \frac \partial {\partial \theta }+i\cos \theta \frac \partial
{\partial \varphi } & 0 \\ 
0 & \sin \theta \frac \partial {\partial \theta }+i\cos \theta \frac
\partial {\partial \varphi }
\end{array}
\right) .  \label{tw20}
\end{equation}

Coming now to the $x$ component, we observe that the eigenvectors for the $x$
component of the spin are[7] 
\begin{equation}
\lbrack \xi _{x,\widehat{{\bf b}}}^{+}]=\frac 1{\sqrt{2}}\left( 
\begin{array}{c}
(\sin \theta /2+\cos \theta /2)e^{-i\varphi /2} \\ 
(\cos \theta /2-\sin \theta /2)e^{i\varphi /2}
\end{array}
\right) \;\;\;\;\;\text{\ }  \label{tw21}
\end{equation}

\begin{equation}
\lbrack \xi _{x,\widehat{{\bf b}}}^{-}]=\frac 1{\sqrt{2}}\left( 
\begin{array}{c}
(\sin \theta /2-\cos \theta /2)e^{-i\varphi /2} \\ 
(\cos \theta /2+\sin \theta /2)e^{i\varphi /2}
\end{array}
\right)  \label{tw22}
\end{equation}
Assuming for the operator $[S_{x,\widehat{{\bf b}}}]$ the form

\begin{equation}
\lbrack S_{x,\widehat{{\bf b}}}]=\left( 
\begin{array}{cc}
A & 0 \\ 
0 & D
\end{array}
\right) ,  \label{tw23}
\end{equation}
with

\begin{equation}
A=a_1\frac \partial {\partial \theta }+a_2\frac \partial {\partial \varphi }
\label{tw24}
\end{equation}
and 
\begin{equation}
D=d_1\frac \partial {\partial \theta }+d_2\frac \partial {\partial \varphi },
\label{tw25}
\end{equation}
we find

\begin{equation}
\lbrack S_{x,\widehat{{\bf b}}}]=\hbar \left( 
\begin{array}{cc}
\cos \theta \frac \partial {\partial \theta }+i\sin \theta \frac \partial
{\partial \varphi } & 0 \\ 
0 & -\cos \theta \frac \partial {\partial \theta }+i\sin \theta \frac
\partial {\partial \varphi }
\end{array}
\right) .  \label{tw26}
\end{equation}

For the $y$ component, the vectors are[7]

\begin{equation}
\lbrack \xi _{y,\widehat{{\bf b}}}^{+}]=\frac 1{\sqrt{2}}\left( 
\begin{array}{c}
e^{-i\varphi /2} \\ 
ie^{i\varphi /2}
\end{array}
\right) \;\;\;\;\;\text{and\ \ \ \ \ \ }[\xi _{y,\widehat{\QTR{textbf}{b}}%
}^{-}]=\frac 1{\sqrt{2}}\left( 
\begin{array}{c}
e^{-i\varphi /2} \\ 
-ie^{i\varphi /2}
\end{array}
\right) .  \label{tw27}
\end{equation}
However, it is found that in order to be able to derive a differential
operator in this case, a phase factor $e^{-i\theta /2}$ must be introduced
into the expressions for the eigenvectors. Thus, we have 
\begin{equation}
\lbrack \xi _{y,\widehat{{\bf b}}}^{+}]=\frac 1{\sqrt{2}}\left( 
\begin{array}{c}
e^{-i(\varphi +\theta )/2} \\ 
ie^{i(\varphi -\theta )/2}
\end{array}
\right) \;\;\;\;\;\text{and\ \ \ \ \ \ }[\xi _{y,\widehat{\QTR{textbf}{b}}%
}^{-}]=\frac 1{\sqrt{2}}\left( 
\begin{array}{c}
e^{-i(\varphi +\theta )/2} \\ 
-ie^{i(\varphi -\theta )/2}
\end{array}
\right) .  \label{tw27a}
\end{equation}
We find 
\begin{equation}
\lbrack S_{y,\widehat{{\bf b}}}]=\hbar \left( 
\begin{array}{cc}
-i\frac \partial {\partial \theta } & 0 \\ 
0 & -i\frac \partial {\partial \theta }
\end{array}
\right) .  \label{tw28}
\end{equation}

The operator for the square of the spin is 
\begin{equation}
\lbrack S^2]=[S_{x,\widehat{{\bf b}}}]^2+[S_{y,\widehat{{\bf b}}}]^2+[S_{z,%
\widehat{{\bf b}}}]^2=\hbar ^2\left( 
\begin{array}{cc}
i\frac \partial {\partial \varphi }-\frac{\partial ^2}{\partial \varphi ^2}
& 0 \\ 
0 & -i\frac \partial {\partial \varphi }-\frac{\partial ^2}{\partial \varphi
^2}
\end{array}
\right)  \label{tw28a}
\end{equation}
The operators we have derived obey the usual commutation relations:

\begin{equation}
\lbrack [S_{i,\widehat{{\bf b}}}],[S_{j,\widehat{{\bf b}}}]]=i\hbar [S_{k,%
\widehat{{\bf b}}}]\;\;\text{with\ \ }i,j,k\text{ in cyclical order,}
\label{tw29}
\end{equation}
and 
\begin{equation}
\lbrack [S_{i,\widehat{{\bf b}}}],[S^2]]=0.  \label{th30}
\end{equation}

\section{Conclusion}

In this paper, we have shown that a differential eigenvalue equation for
spin 1/2 exists. We would hope to extend our arguments to other values of
spin. Unlike the case of orbital angular momentum, it is clear that there is
not one differential spin operator but as many as there are different values
of spin. It would be surprising if differential eigenvalue equations did not
exist for other values of spin. A very important question to be answered is
the interpretation of the differential operator. This may conceivably lead
to a clarification of the nature of intrinsic spin.

As the eigenfunctions satisfying our eigenvalue equation refer to the $z$
direction as the initial direction of quantization, we have not obtained the
most general form of the differential eigenvalue equation for spin 1/2. Such
a form would have an arbitrary direction as the initial direction of
quantization. The search is proceeding for this general form.

\section{References}

1. Land\'e A., ''From Dualism To Unity in Quantum Physics'', Cambridge
University Press, 1960.

2. Land\'e A., ''New Foundations of Quantum Mechanics'', Cambridge
University Press, 1965.

3. Land\'e A., ''Foundations of Quantum Theory,'' Yale University Press,
1955.

4. Land\'e A., ''Quantum Mechanics in a New Key,'' Exposition Press, 1973.

5. Mweene H. V., ''Derivation of Spin Vectors and Operators From First
Principles'', quant-ph/9905012

6. Mweene H. V., ''New Treatment of Systems of Compounded Angular
Momentum'', quant-ph/9907082

7. Mweene H. V., ''Generalized Spin-1/2 Operators and Their Eigenvectors'',
quant-ph/9906002

\end{document}